\documentclass[aps,prx,superscriptaddress,amsmath,amssymb,reprint,showkeys]{revtex4-2}

\usepackage{graphicx}
\usepackage{xcolor}
\usepackage{textcomp}
\usepackage{booktabs}

\usepackage[separate-uncertainty=true]{siunitx}
\usepackage{enumitem}
\usepackage[version=4]{mhchem}
\usepackage{upgreek}
\usepackage[breaklinks=true,colorlinks=true,linkcolor=blue,urlcolor=blue,citecolor=blue]{hyperref}

\makeatletter
\def\@email#1#2{%
	\endgroup
	\patchcmd{\titleblock@produce}
	{\frontmatter@RRAPformat}
	{\frontmatter@RRAPformat{\produce@RRAP{*#1\href{mailto:#2}{#2}}}\frontmatter@RRAPformat}
	{}{}
}%
\makeatother

\newcommand{\upr}[2]{#1_{\mathrm{#2}}}

\begin{document}

\title{An all-magnonic neuron with tunable fading memory} 

\author{David Breitbach}
\email{dbreitba@rptu.de}
\altaffiliation{D. Breitbach and M. Bechberger contributed equally to this work.}
\affiliation{Fachbereich Physik and Landesforschungszentrum OPTIMAS, RPTU Kaiserslautern-Landau, \mbox{D-67663 Kaiserslautern, Germany}}

\author{Moritz Bechberger}%
\altaffiliation{D. Breitbach and M. Bechberger contributed equally to this work.}
\affiliation{Fachbereich Physik and Landesforschungszentrum OPTIMAS, RPTU Kaiserslautern-Landau, \mbox{D-67663 Kaiserslautern, Germany}}

\author{Hanadi Mortada}
\affiliation{Fachbereich Physik and Landesforschungszentrum OPTIMAS, RPTU Kaiserslautern-Landau, \mbox{D-67663 Kaiserslautern, Germany}}

\author{Bj{\"o}rn Heinz}
\affiliation{Fachbereich Physik and Landesforschungszentrum OPTIMAS, RPTU Kaiserslautern-Landau, \mbox{D-67663 Kaiserslautern, Germany}}

\author{Roman Verba}
\affiliation{\mbox{V. G. Baryakhtar Institute of Magnetism of the NAS of Ukraine, Kyiv 03142, Ukraine}}

\author{Qi Wang}
\affiliation{{School of Physics, Hubei Key Laboratory of Gravitation and Quantum Physics, Institute for Quantum Science and Engineering, Huazhong University of Science and Technology, Wuhan, China}}

\author{\mbox{Carsten Dubs}}
\affiliation{\mbox{INNOVENT e.V. Technologieentwicklung, D-07745 Jena, Germany}}

\author{Mario Carpentieri}
\affiliation{Department of Electrical and Information Engineering, Politecnico di Bari, \mbox{70125 Bari, Italy}}

\author{Giovanni Finocchio}
\affiliation{Department of Mathematical and Computer Sciences, Physical Sciences and Earth Sciences, University of Messina, \mbox{98166 Messina, Italy}}

\author{Davi Rodrigues}
\affiliation{Department of Electrical and Information Engineering, Politecnico di Bari, \mbox{70125 Bari, Italy}}

\author{Alexandre Abbass Hamadeh}
\affiliation{Centre de Nanosciences et de Nanotechnologies, Universit\'e Paris-Saclay, \mbox{91120 Palaiseau, France}}

\author{Philipp Pirro}
\affiliation{Fachbereich Physik and Landesforschungszentrum OPTIMAS, RPTU Kaiserslautern-Landau, \mbox{D-67663 Kaiserslautern, Germany}}

\begin{abstract}
Magnonics offers nanometer-scale wave propagation and strong nonlinearities, making it attractive for neuromorphic applications such as artificial neurons. Yet, magnonic elements with interconnections solely within the magnonic system remain challenging, preventing the realization of interconnected magnonic neurons to date. Here, we experimentally demonstrate an all-magnonic neuron that reacts to magnon inputs with thresholded, amplified magnon firing and subsequent self-reset, enabling all-magnonic operation and cascading. Our approach is based on micro-antenna excitation on an ultra-low damping garnet with perpendicular magnetic anisotropy (PMA), where we exploit the positive magnon frequency shift to realize nonlinear activation. Using Brillouin light scattering spectroscopy, we uncover a transient neuron response with tunable fading memory: A 25\% change in pump power results in a 3-order-of-magnitude tuning in memory time, which we harness, demonstrating temporal integration of up to 50 magnon pulses. Finally, we realize neuron triggering in a cascade of 3 neurons, highlighting its potential for connected magnonic circuits.
\end{abstract}


\maketitle

\clearpage


\section*{Introduction}\label{sec:intro}

The advancement of artificial neural networks has created a great research interest in artificial, analog neurons \cite{markovic.2020.physics, kudithipudi.2025.neuromorphiccomputingscale}. Specifically, analog systems relying on wave-based phenomena are interesting for their potential to combine parallel processing capabilities, intrinsic time-domain computation and connectivity for neuromorphic approaches \cite{Hughes.2019.recurrentneural, lin.2018.diffractive-deep-NNs,zuo.2019.All-optical-NN-nonlinearAct, rodrigues.2023.dynamic-spintronic-NN, grollier.2020.neuromorphic-spintronics, ross.2023.multilayer-spintronic-NNs}. Wave interference, for example, is promising for the parallel execution of a weighted summation, and other core elements of neural computation \cite{katayama.2016.wave-based-computing, rahman.2015.wave-inteference-functions}. Furthermore, wave-based systems enable 2D wave propagation and planar waveguide crossings via waveguide coupling \cite{papp.2021.spinwaveNN, wang.2018.directional-coupler}, offering a potential solution to the increasing wiring problem in neural network hardware architectures. 

Among these approaches, magnons—the quanta of spin waves—offer a promising platform for neuromorphic hardware \cite{pirro.2021.advancescoherentmagnonics, chumak.2022.roadmap-SW-computing, csaba.2017.perspectives-SW-for-computing, wang-2024.nanoscale-magnonic-networks}. Magnons exhibit strong intrinsic nonlinearities \cite{an.2024.emergentcoherentmodes} crucial for neuromorphic functions \cite{flebus.2024.magnonics-roadmap-2024, papp.2021.spinwaveNN}. They operate naturally in the gigahertz frequency range and exhibit wavelengths down to the nanometer scale. Furthermore, magnons can be reconfigured by changing the magnetic ground state, perspectively enabling synaptic weights via micromagnets \cite{concconcelli.2024.nanomagnets}, domain walls \cite{wagner.2016.domainwall-reconfig-SW-nanochannels, fan.2023.domain-wall-motion-by-SW} or by local tuning of the magnetic anisotropy \cite{levati.2025.threedimensionalnanoscalecontrola, chen.2025.VCMA-half-adder}. 

Magnonic systems have led to the design of several neuromorphic systems \cite{ustinov.2024.magnonic-reservoir, watt.2021.magnonic-reservoir-time-delay, papp.2021.spinwaveNN, korber.2023.patternrecognitionreciprocal, gartside.2022.reconfigurabletrainingreservoir, nishimura.2026.highaccuracypredictionnarma}, yet, the realization of magnonic neurons has not been achieved to date. A main obstacle has been the implementation of magnonic elements that can be triggered via magnons and actively emit amplified magnon signals upon activation, enabling their all-magnonic interconnectivity. First theoretical studies on magnonic neuron concepts have shown great potential to realize neuron-like nonlinear activation using magnonic resonators \cite{kruglyak.2021.chiral-resonators, fripp.2023.toward-magnonic-neurons}. As passive transmission devices without gain, however, these concepts rely on means of external amplification. This desirable functionality was recently shown in repeater-type devices \cite{wang.2024.repeater}. However, their response resembles a one-way switch which erases any temporal information from the input signal, and lacks a self-reset mechanism, i.e. remaining active once turned on.

In this study, we combine the advantages of both approaches: We experimentally realize an all-magnonic neuron which intrinsically supports inter-neuron connectivity via magnon triggering and amplified emission, and shows autonomous decay ('self-reset') after triggering. Specifically, our device integrates both, transient nonlinear response and magnon amplification in a single physical element. This enables interconnected neuron cascades, as well as temporal processing functions such as leaky integration and fading memory, and eliminates the need for an external clock. To understand the neuron mechanism, we study its function in four parts:

\begin{enumerate}[label=(\roman*),itemsep=1mm, parsep=0pt]
    \item A nonlinear excitation mechanism for propagating magnons
    \item Triggering of this mechanism via external magnon pulses
    \item Temporal integration of multiple temporal inputs
    \item Cascaded triggering of multiple neurons
\end{enumerate}

In the following, we will show a nonlinear excitation mechanism for propagating magnons (i) which is demonstrated via direct magnon excitation. As schematically displayed in Fig. \ref{fig:device_layout}a, we then apply externally excited, propagating magnonic inputs to the neuron and study its triggering and nonlinear response (ii). Our findings reveal a transient neuron activation in the form of propagating magnon emission, which we then apply to study the temporal integration of multiple input pulses (iii) and cascaded triggering of three neurons (iv). Finally, we use a simplified neural network model to verify whether the experimentally observed activation nonlinearity is already sufficient for standard learning tasks.

\section*{Results}\label{sec:results}
\subsection*{Device Layout}\label{sec:device}

\begin{figure}[ht!]
\centering
\includegraphics[]{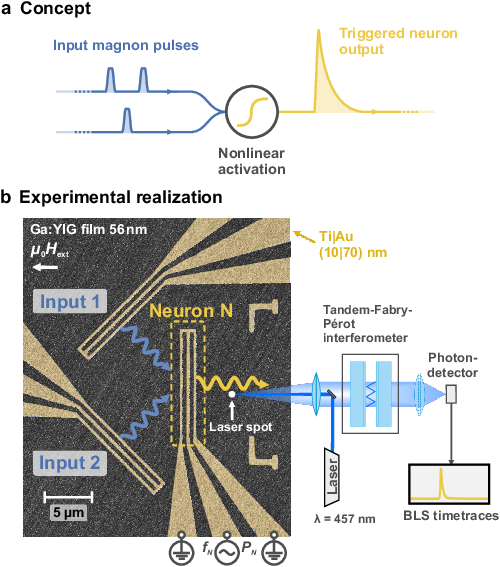}
\caption{\textbar\enspace\textbf{Device layout}. \textbf{a} Conceptual diagram of the neuron principle: The neuron is triggered when incoming magnon stimuli exceed the threshold given by its nonlinear activation function, leading to the emission of a magnon pulse, before self-resetting back to baseline. \textbf{b} Colorized SEM micrograph of the applied microstructure, consisting of a vertical CPW antenna, here functionally representing the neuron, and two diagonally placed CPW antennas used for input generation. The structures are placed on a Ga:YIG film, which is magnetized in-plane using a bias magnetic field of $\mu_{\mathrm{0}}H_{\mathrm{app}} \approx \SI{86}{\milli\tesla}$ along the indicated direction. The emitted magnons are measured using a time-resolved BLS microscope, which is schematically depicted on the right. }\label{fig:device_layout}
\end{figure}

\begin{figure*}[t!]
\centering
\includegraphics[width = \textwidth]{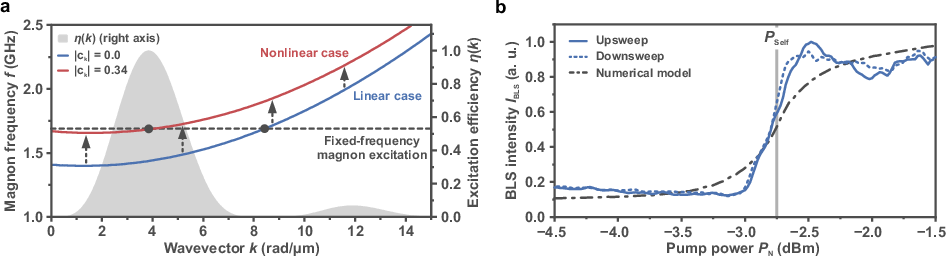}
\caption{\textbar\enspace\textbf{Nonlinear excitation mechanism.} \textbf{a} Left axis: Magnon dispersion relation $\upr{f}{k} = \omega_{\mathrm{k}}/2\pi$ for vanishing amplitude $c_{\mathrm{k}} = 0.0$ (blue) and with nonlinear magnonic self- and cross-frequency shift for increased amplitude $|c_{\mathrm{k}}| = 0.34$ (red) of mode $k \approx \SI{8.3}{\radian\per\micro\meter}$. Right axis: CPW excitation efficiency $\eta(k)$ as a function of the magnon wavevector $k$ (gray). \textbf{b} Nonlinear dependence of the BLS intensity $I_{\mathrm{BLS}}$ of the emitted magnons as a function of up- (blue, continuous) and downsweep (blue, dashed) of the excitation power $P_{\mathrm{N}}$. The results of the numerical model are shown for comparison (black, dashed).}
\label{fig:excitation}
\end{figure*}

Experimentally, our approach is based on a gallium-substituted yttrium iron garnet (Ga:YIG) thin film ($t = \SI{56}{\nano\meter}$) \cite{dubs.2025.gayig} with three nanofabricated coplanar waveguide (CPW) antennas on top, see Fig. \ref{fig:device_layout}b. Principally, each of these antennas can take over the function of a neuron. In the following, the vertical antenna is used as a neuron, referred to by the index N, while the diagonal antennas serve as the (magnonic) inputs for single or multiple input signals with different timings, here referred to as input 1 and 2, respectively. We later study the neurons response to these magnonic input stimuli using a BLS microscope. To realize this response, however, we first establish the nonlinear excitation mechanism \textit{without triggering} in the next section.

\subsection*{Nonlinear activation mechanism}\label{sec:mechanism}
Magnon excitation is induced via dynamic Oersted fields by applying a radio frequency (RF) current to the neuron antenna. The idea is to realize the nonlinear activation function by relying on the nonlinear frequency shift of the magnon dispersion relation with the magnon intensity, which is displayed in Fig. \ref{fig:excitation}a. This frequency shift is caused by the four-magnon interaction of the excited mode $\mathbf{k}$ with itself and/or its cross-interaction with other modes $\mathbf{k'}$. The displayed nonlinear dispersion relation $\omega_{\mathbf{k}}$ is calculated via \cite{lvov.1994.nonlinearshift}:
\begin{equation}\label{equ:nonlinear_shift}
        \tilde\omega_\mathbf{k} = \omega_\mathbf{k} + T_\mathbf{k} |c_\mathbf{k}|^2 + 2 T_{\mathbf{k}\mathbf{k'}} |c_\mathbf{k'}|^2 \,,
\end{equation}\\
with the nonlinear frequency self- and cross-shift coefficients $T_\mathbf{k}$ and $T_\mathbf{kk'}$, following the Hamiltonian formalism for nonlinear magnon dynamics \cite{patton.2010.hamiltonian} (Methods and Supplementary Fig. S1). Importantly, the frequency shift depends on the magnon intensity and, therefore, on the RF power applied to the CPW.

Generally, a driven oscillator with a power-dependent resonance frequency exhibits a foldover effect---a modification of the resonance curve which, under certain conditions, causes threshold-like behavior and/or bistability, as the frequency of the system shifts 'into resonance' with the driving source. This foldover effect is well known from Duffing-type nonlinear oscillators \cite{Rabinovich.1989.Book} and has been observed in various magnonic systems with discrete (standing modes) or continuous (propagating magnons) spectra \cite{gui.2009.foldover, janantha.2017.foldover,chen.1989.FMR-foldover, wang.2020.ringresonator}. 

We functionalize this effect by using Ga:YIG, which shows a large PMA and low $\upr{M}{s}$, resulting in a \textit{positive} nonlinear frequency shift for in-plane magnetization \cite{breitbach.2024.erasing}. As a result, by excitation with a CPW antenna at a frequency \textit{above} the linear resonance frequency, the antenna excitation efficiency is strongly enhanced with the magnon intensity (see Fig.~\ref{fig:excitation}a), resulting in a positive feedback mechanism. This synergy has been used in prior works for the excitation of short-wavelength magnons and other functionalities \cite{wang.2023.oopyig, wang.2024.fastswitchable}. Most importantly, the positive shift enables the resonant emission of propagating magnons above the adjacent material band-bottom, i.e., neuron firing. This is in contrast to commonly used systems with negative frequency shifts, where nonlinear self-localization effects hinder deep nonlinear effects relying on propagating magnons \cite{Slavi.2005.SW-Bullet, Sulymenko.2018.SW-Solitons}

Hence, this positive feedback reflects in the emitted magnon BLS intensity $I_{\mathrm{BLS}}$, which becomes a nonlinear function of the applied power, see Figure \ref{fig:excitation}b. While the excited magnon intensity is almost constant for high and low powers, an increase by a factor of $5$ is observed in a small power window of about $\SI{0.4}{dB}$. This behavior is the basis for the nonlinear activation function of the neuron. For the particular neuron, we can define the self-activation power by $P_{\mathrm{Self}} = \SI{-2.75}{dBm}$, and subsequent power values are given in relative terms $P'_{\mathrm{N}} = P_{\mathrm{N}} - P_{\mathrm{Self}}$. 

As seen in the up- and downsweep curves, this extended film exhibits a continuous response without hysteresis. We note that the foldover effect can in general result in both hysteretic (bistable) \cite{wang.2023.oopyig,wang.2024.repeater} and non-hysteretic power-dependencies, which is primarily determined by the excitation frequency, but also sensitive to local inhomogeneities, anisotropies, thermal, and other effects. Later, we demonstrate how bistable behavior can enable additional neuron functionalities.

To understand how we can externally trigger the neuron, we model the above process using a simplified flux rate approach by a first-order nonlinear ordinary differential equation describing the temporal evolution of the magnon intensity:
\begin{equation}\label{equ:ODE}
\begin{split}
\frac{\mathrm{d}|\upr{c}{N}(t)|^2}{\mathrm{d}t}
    &= \eta(k[|\upr{c}{total}(t)|^2])\upr{P}{MW} \\
    &\quad - \lambda|\upr{c}{N}(t)|^2
      - \lambda'|\upr{c}{N}(t)|^4.
\end{split}
\end{equation}

\begin{figure*}[t!]
\centering
\includegraphics[width=\textwidth]{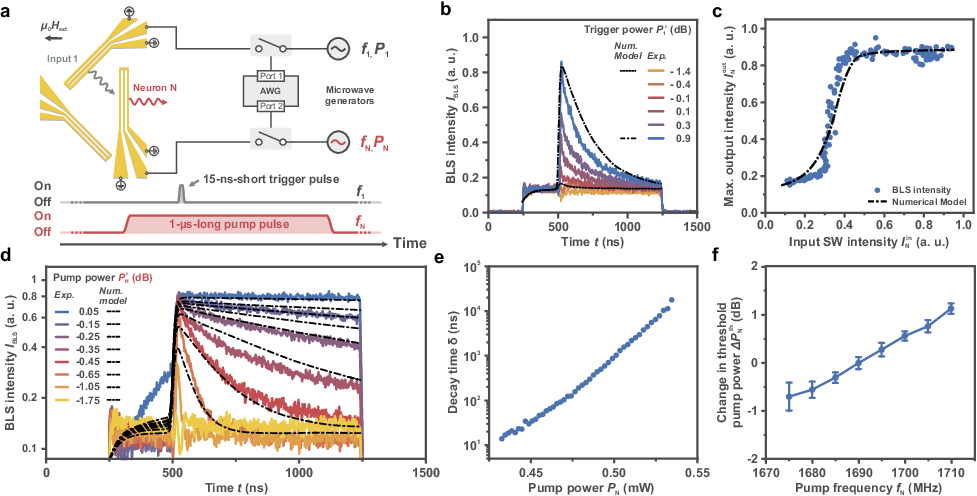}
\caption{\textbar\enspace\textbf{Triggered neuron activation and decay.} \textbf{a} Experimental schematic: Two RF-pulses of duration $\Delta\tau_{\mathrm{1}} = \SI{15}{\nano\second}$ and $\Delta\tau_{\mathrm{N}} = \SI{1}{\micro\second}$ and frequency $f_{\mathrm{1}} = \SI{2.2}{\giga\hertz}$ and $f_{\mathrm{N}} = \SI{1.69}{\giga\hertz}$ are generated and applied to input 1 and to the neuron, respectively. Note that $f_{\mathrm{1}}$ was optimized for high pulse quality (low pulse dispersion, high $\upr{v}{G}$), but triggering with other frequencies is generally possible if enough positive nonlinear frequency shift can be induced, see also Fig. \ref{fig:cascade}. \textbf{b} Temporal evolution of the neuron output BLS intensity and numerical model predictions for fixed pump power $P_{\mathrm{N}}' = \SI{-0.45}{dB}$ and different trigger powers $P_{\mathrm{1}}' = P_{\mathrm{1}} - P_{\mathrm{Trigger}}'$, where $P_{\mathrm{Trigger}}$ is the absolute trigger power yielding 50\% activation (with regard to intensity maximum). \textbf{c} Neuron activation function: maximum neuron output BLS intensity vs maximum input BLS intensity by input 1 and numerical model for comparison. \textbf{d} Tunable neuron decay: Variation of the pump power $P_{\mathrm{N}}'$ for a fixed trigger power $P_{\mathrm{1}}' = \SI{1.1}{dB}$. \textbf{e} Neuron decay time extracted from exponential decay fits from panel (d) as a function of the pump power $P_{\mathrm{N}}$ in mW. \textbf{f} Change of threshold pump power $P_{\mathrm{N}}^{\mathrm{th}}$ for triggered activation at a fixed trigger power $P_{\mathrm{1}}' = \SI{1.1}{dB}$ as a function of the pump frequency $f_{\mathrm{N}}$.} \label{fig:activation}
\end{figure*}

The first term on the right-hand side describes the direct RF-excitation with efficiency $\eta(k)$ and power $P_{\mathrm{MW}}$. The total magnon intensity dynamically affects the wavevector with two contributions $|\upr{c}{total}(t)|^2 = |\upr{c}{N}(t)|^2 + |\upr{c}{in}(t)|^2$: magnons excited by the neuron itself and \textit{inputs from external sources} such as other neurons, both driving a nonlinear frequency shift via Eq. (\ref{equ:nonlinear_shift}). The last two terms in Eq. (\ref{equ:ODE}) phenomenologically model the magnon loss by linear and nonlinear damping as is common for high-intensity magnonic systems \cite{slavin.2009.nonlinear-SAO, scott.2004.nonlineardampinghighpower}. Equation \ref{equ:ODE} is solved with temporal intensity noise and statistical averaging to improve comparability with stroboscopic measurements and to avoid metastable solutions, which are unlikely in our experiment due to the spatial inhomogeneity of the extended magnetic medium. Details about the numerical model can be found in the Methods and Supplementary Fig. S2.

In the simplest case of self-activation, $|c_{\mathrm{in}}(t)|^2 = 0$, the neuron is driven purely by its own RF-excitation. This stationary case, shown in Fig. \ref{fig:excitation}b, yields a nonlinear intensity-power relation that agrees qualitatively with the experimental data. However, a propagating magnon pulse can also induce a large nonlinear frequency shift even at significant distances from its source \cite{breitbach.2024.erasing}. Motivated by this nonlocal interaction, we now consider the intriguing scenario in which a neuron is triggered by magnons originating from external inputs, i.e., $|c_{\mathrm{in}}(t)|^2 > 0$, highlighting how the same nonlinear mechanism could later enable interaction between neurons.

\subsection*{Neuron triggering}\label{sec:trigg}
We demonstrate external triggering of the neuron by exploiting the nonlinear frequency shift induced by incoming magnons from the adjacent input 1. The corresponding setup is illustrated in Fig. \ref{fig:activation}a: The neuron is driven with a 1-µs-long RF \textit{pump pulse} with a pump power just below its self-activation point ($P_{\mathrm{N}}' < 0$). During this pump pulse, a 15-ns-short magnon \textit{trigger pulse}, emitted from input 1, delivers enough nonlinear shift to bring the neuron temporarily above the nonlinear threshold, i.e., in a state of much higher excitation efficiency $\eta$. The triggered response of the neuron is shown in Fig. \ref{fig:activation}b for several trigger powers $P_{\mathrm{1}}$. At the beginning of the pump pulse, the neuron's sub-threshold excitation can be observed, followed by an almost instantaneous intensity increase at $t = \SI{500}{\nano\second}$ when the trigger pulse reaches the neuron, an event we refer to as \textit{neuron activation}. Since the pump power is below the self-activation power $P_{\mathrm{N}} < P_{\mathrm{Self}}$, the neuron cannot stabilize this high level of intensity after the short input pulse, and decays back to its pre-activation level. In the numerical model, this external trigger pulse is taken into account by modelling $|c_{\mathrm{in}}(t)|^2$ as a time-dependent input pulse that drives a cross-mode nonlinear frequency shift and thereby increases $\eta$. The resulting solutions, represented by the black dashed lines, qualitatively reproduce the experimental findings and are shown exemplarily for the lowest and highest power levels.

\begin{figure*}[t!]
\centering
\includegraphics[width = \textwidth]{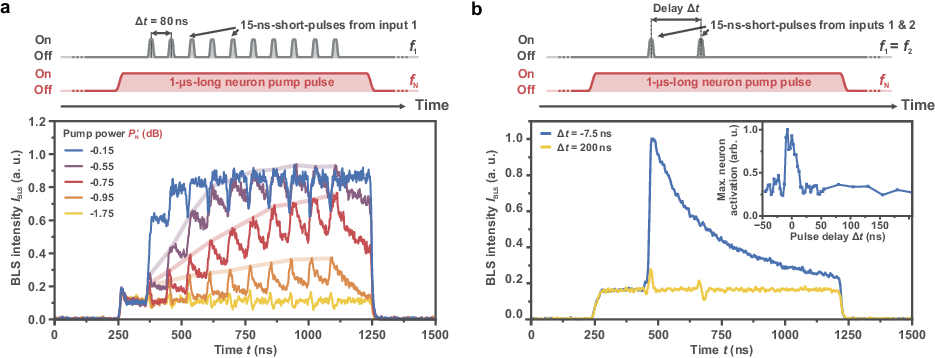}
\caption{\textbar\enspace\textbf{Multi-input neuron triggering.} \textbf{a} BLS intensity of the neuron, triggered by 10 consecutive 15-ns-long input pulses from input 1, each with a pulse spacing of $\Delta t = \SI{80}{\nano\second}$ and frequency $f_{\mathrm{1}} = \SI{2.2}{\giga\hertz}$. Shown are different curves for varying pump powers $P_{\mathrm{N}}'$, at the pump frequency $f_{\mathrm{N}} = \SI{1.69}{\giga\hertz}$. The peak positions are indicated by a semi-transparent line as a guide to the eye. \textbf{b} BLS intensity of the neuron, triggered by both inputs 1 and 2, each sending an independent 15-ns-short trigger pulse with varying pulse delay between them, with frequencies $f_{\mathrm{1}} = f_{\mathrm{2}} = \SI{2.2}{\giga\hertz}$. Measured at the output side of the neuron with a pump power of $P_\mathrm{N}' = \SI{-0.7}{dB}$ and pump frequency $f_{\mathrm{N}} = \SI{1.69}{\giga\hertz}$. \textbf{Inset}: Maximum neuron output as a function of the pulse delay $\Delta t_{\mathrm{1,2}}$ between the two pulses.}\label{fig:multi}
\end{figure*}

The intensity of the activated neuron output depends on the input trigger power. To determine the neuron activation function, its maximum intensity is extracted, and the input trigger pulse intensity is additionally measured at the same distance in front of the neuron. Fig. \ref{fig:activation}c shows the resulting neuron output vs. input function. For low input intensities, the neuron operates at its sub-threshold excitation level. For higher input intensities, a nonlinear step is observed before the neuron output saturates at even higher intensities.

After activation, the neuron's intensity decays autonomously, resetting its state without requiring an external clock. This event-driven behavior mirrors biological neurons, leaving the neuron inactive when the input is below the activation threshold, which contributes to the energy efficiency of neuromorphic systems. While faster decay means higher clock speed, the decay process also provides a form of volatile \textit{fading memory}: an echo of recent activity that allows the neuron to respond based on the temporal structure of its inputs \cite{maass.2002.LSM-fading-memory, maass.2004.computationalpowercircuits, roy.2019.spikebasedmachineintelligence}. This makes operations like leaky integration naturally accessible within the system.

Importantly, in our system, the decay time is not fixed, but tunable. As shown in Fig. \ref{fig:activation}d, varying the pump power of $P_{\mathrm{N}}$ while keeping the trigger power constant allows control over the relaxation dynamics. As the pump power approaches the self-activation power $P_{\mathrm{Self}}$, the neuron is brought closer to a state where it can sustain activation on its own, and the decay time increases significantly. When the power $P_{\mathrm{Self}}$ is exceeded (blue curve), the neuron self-activates even before the trigger and stabilizes the high-intensity state indefinitely. Notably, we observe decay times up to several microseconds, exceeding the intrinsic magnon lifetime by orders of magnitude (Supplementary Fig. S3). This extended memory window arises from the interplay of the pump-driven inflow, and the damping-driven outflow of magnons, in combination with the nonlinear coupling to the pump. Once activated, the system remains in a state of increased excitation efficiency for a time that depends sensitively on its proximity to the nonlinear threshold. The model of Eqn. (\ref{equ:ODE}) captures the trend qualitatively: higher pump power leads to longer decay.

Figure \ref{fig:activation}e shows the extracted decay times from exponential decay fits as a function of pump power (linear units). A $\sim\SI{25}{\percent}$ variation of power results in a three-order-of-magnitude change in decay time, demonstrating dynamic tunability over a wide timescale. A change in the input power, on the other hand, leaves the decay times almost unchanged (Supplementary Fig. S4). Beyond decay dynamics, the nonlinear threshold itself can also be tuned by varying the excitation frequency, as shown in Fig. \ref{fig:activation}f. Together, pump power and frequency provide two independent parameters to control the neuron's temporal behavior and activation characteristics.\\
\newline

\subsection*{Multi-input neuron triggering}\label{sec:multi}

The neuron's fading memory enables temporal integration of subsequent inputs over time. This is demonstrated by sending a train of ten 15-ns-long pulses spaced by $\Delta t = \SI{80}{\nano\second}$ from input 1 to the neuron, as shown in Fig. \ref{fig:multi}a. The pump power $P_{\mathrm{N}}$ is varied to explore different regimes of decay and activation. At low pump powers, the neuron exhibits a weak response to each individual pulse. As the pump power increases, both the immediate excitation and decay time grow, allowing residual excitation from previous pulses to accumulate. This results in an incremental neuron activation with each successive pulse. At intermediate powers, a linear increase of the local maxima is observed (shaded line), effectively implementing a leaky integrator where the maxima in the output are directly proportional to the number of input pulses, compare also Ref. \cite{bracher.2018.magnon-adder}. This functionality was demonstrated with as many as 50 pulses over \SI{2}{\micro\second} (Supplementary Fig. S5). At higher pump powers, the accumulation becomes nonlinear, and saturation occurs toward the end of the pulse train.

In addition to temporal integration, the neuron can be triggered by the combined input from multiple, spatially separated sources. In the experiment shown in Fig. \ref{fig:multi}b, two 15-ns-long trigger pulses are emitted from the inputs 1 and 2 toward the neuron with a tunable time delay $\Delta t$. Contrary to the previous case, the pump and trigger powers are chosen such that either pulse alone is insufficient to activate the neuron. Hence, for large delays ($\Delta t = \SI{200}{\nano\second}$), only small, sub-threshold responses are seen. However, when the pulses overlap in time (blue curve), the induced nonlinear shift momentarily exceeds the critical threshold, resulting in neuron activation. The inset of Fig. \ref{fig:multi}b shows a coincidence window of approximately $\SI{30}{\nano\second}$. This is a basic realization of a two-input-neuron acting as a time-domain AND gate with autonomous decay. More importantly, it provides a first experimental indication that a single magnonic neuron can be driven by multiple presynaptic inputs.  

\subsection*{Cascaded Neuron Activation}\label{sec:Cascade}

\begin{figure}[t!]
\centering
\includegraphics[width=0.45\textwidth]{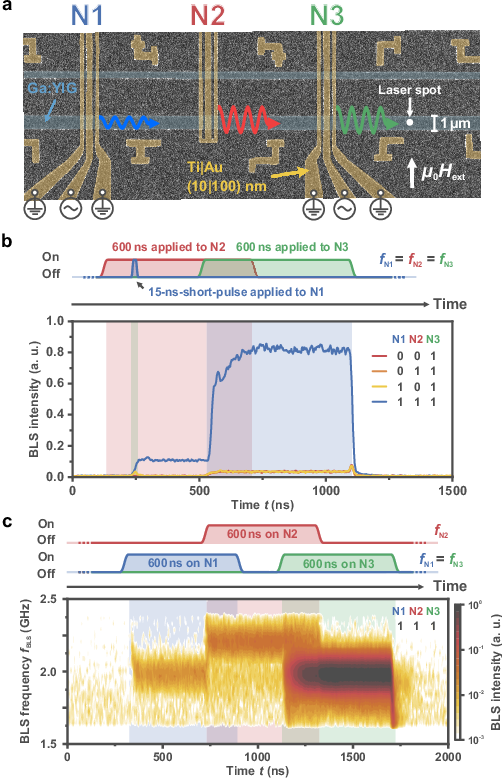}
\caption{\textbar\enspace\textbf{Cascaded neuron activation.} \textbf{a} Colorized SEM micrograph of the employed nanostructure: Three CPW antennas are placed on top of a 1-$\upmu$m-wide Ga:YIG waveguide. The neurons N2 and N3 are driven at powers in the bistability power regime $\upr{P}{N2} = \SI{-2}{dBm}$ (compare Supplementary Fig. S6). BLS measurement is performed on the right side of N3, 'behind' the chain of neurons. \textbf{b} Cascaded neuron experiment, with one trigger pulse by N1 and two pump pulses by N2 and N3, showing the time-resolved BLS intensity for combinations of pump states (ON/OFF) (binarized in legend). The pulse timings are indicated by the shaded areas. Frequencies used: $f_{\mathrm{N1}} = f_{\mathrm{N2}} = f_{\mathrm{N3}} = \SI{2.1}{\giga\hertz}$. \textbf{c} Repetition of a similar experiment with a different pump frequency for the center neuron $f_{\mathrm{N2}} = \SI{2.2}{\giga\hertz}$, and $f_{\mathrm{N1}} = f_{\mathrm{N3}} = \SI{2.0}{\giga\hertz}$ yielding similar results as (b). Shown is the BLS intensity as a function of time and BLS frequency for the (1 1 1) case. }\label{fig:cascade}
\end{figure}

To go beyond single-neuron operation, we now consider a linear arrangement of three antennas, in which each antenna is regarded as an individual neuron. In this geometry, magnons emitted by one neuron propagate along a 1-$\upmu$m-wide Ga:YIG waveguide and can trigger the next neuron in sequence, enabling a directed cascaded activation chain (N1 $\rightarrow$ N2 $\rightarrow$ N3) mediated by magnon transmission between neighboring elements, see Fig.~\ref{fig:cascade}a. In the present experiment, we use the first antenna to launch the initial magnon pulse. We denote it as neuron N1 to emphasize that it represents the same neuron element as N2 and N3 (as established in the preceding sections) and could, in principle, also be triggered by magnons from an upstream source. The focus of this section is the cascadability motif, i.e., that propagating magnons emitted by one neuron can activate the next neuron in sequence.

In the confined waveguide, the nonlinear excitation exhibits a bistable power window for the selected excitation frequency (Supplementary Fig.~S6). This provides an additional operational mode in which a neuron can persist in a high-output state once it has been triggered. We emphasize that operation at pump powers \textit{below} the bistable window still yields the autonomously decaying, fading-memory neuron behavior discussed in the previous sections, as shown in Supplementary Fig.~S6. Here, however, we intentionally use the bistable regime to simplify the interpretation and highlight the cascade in a particularly robust form.

A further characteristic of this waveguide geometry is strongly non-reciprocal emission, a typical feature of the Damon--Eshbach configuration due to the chirality of magnetization precession and the excitation field \cite{schneider.2008.efficiency} (Supplementary Fig.~S7). This results in predominantly forward emission along the waveguide, supporting a directed, feed-forward cascade.

For the cascaded activation experiment, neurons N2 and N3 are driven by 600-ns-long RF pump pulses within their bistable regime. Although the applied pump powers are above the nonlinear threshold, both neurons remain in their low-intensity state in the absence of a trigger. The cascade is initiated by a 15-ns trigger pulse applied to neuron N1. In the first implementation, all three neurons are operated at the same frequency, $f_{\mathrm{N1}} = f_{\mathrm{N2}} = f_{\mathrm{N3}} = \SI{2.1}{\giga\hertz}$. Pulse timings are chosen to separate the individual activation steps and to distinguish a sequential cascade (N1 $\rightarrow$ N2 $\rightarrow$ N3) from a combined triggering of the last neuron by simultaneous activity of the first two neurons (N1 + N2 $\rightarrow$ N3). The states of the pumps (ON/OFF) are indicated in binarized format in the legend in  Fig.~\ref{fig:cascade}b. The resulting time-resolved $\upmu$BLS intensity shows that neuron N3 remains inactive when either only N1 or only N2 is pumped, and that N3 is activated only when both preceding neurons participate in the chain. In the activated case, the output intensity of N3 increases by approximately a factor of 25 compared to its sub-threshold excitation level (Fig.~\ref{fig:cascade}b), demonstrating a strong nonlinear response of the cascaded neuron.

We also perform a variant of this experiment using distinct excitation frequencies for the center neuron, $f_{\mathrm{N2}} \neq f_{\mathrm{N1}} = f_{\mathrm{N3}}$, while maintaining otherwise similar conditions (Fig.~\ref{fig:cascade}c). The time and frequency resolution of $\upmu$BLS allows us to distinguish the different signals. Again, the last neuron is activated only when the preceding neurons are active, even though the center neuron differs in frequency by $\Delta f \approx \SI{200}{\mega\hertz}$. This behavior is consistent with the fact that the nonlinear self- and cross-shift coefficients remain positive over a broad wavevector range (Supplementary Fig.~S1), so that magnons at different frequencies can still provide the nonlinear shift required for activation. This illustrates that the triggering mechanism is not restricted to strictly identical frequencies, but can also operate in a frequency-mismatched setting.

\begin{figure*}[t!]
\centering
\includegraphics[width=\textwidth]{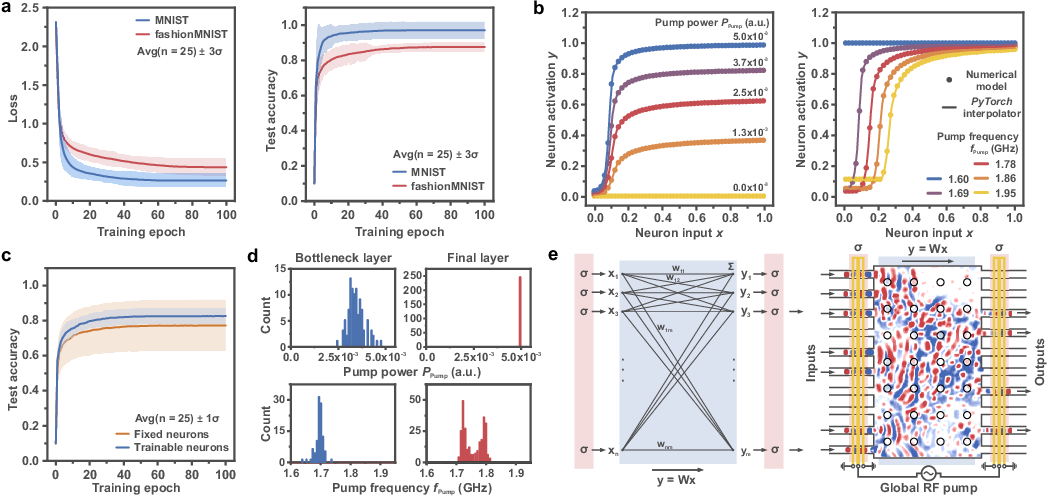}
\caption{\textbar\enspace\textbf{Training results, trainable activation, and a potential route to a magnonic neural network.} \textbf{a} Training loss and test accuracy of the convolutional neural network on \textit{MNIST} and \textit{fashionMNIST}, averaged over $n = 25$ runs with varying parameter initializations. \textbf{b} Single neuron activation function predicted by the numerical model as a function of $P_{\mathrm{Pump}}$ and $f_{\mathrm{Pump}}$ and the corresponding \textit{PyTorch} interpolator used for training. \textbf{c} Comparison of the test accuracy of the bottleneck CNN for training with fixed vs trainable activation functions on average across $n=25$ training runs. \textbf{d} Distribution of the $f_{\mathrm{Pump}}$ and $P_{\mathrm{Pump}}$ parameters after training for the second to last layer (bottleneck) and final layer. The initialization of these parameters was constant at $f_{\mathrm{Pump}} = \SI{1.7}{\giga\hertz}$ and $P_{\mathrm{Pump}}=\SI{2.5e-3}{}$ to be comparable to the fixed case. \textbf{e} Schematic illustration of a potential path towards a fully connected magnonic layer between $10 \times 10$ of the proposed neurons using reconfigurable spin-wave scatterers as proposed in Ref. \cite{papp.2021.spinwaveNN} and schematic representation of the corresponding connection layout.} \label{fig:neuralnetwork}
\end{figure*}

Note that bistability based operation implies that an external reset or clocking scheme is required to return neurons to the low-output state once triggered. At present, a general mechanism to reliably reset a locally triggered bistable (foldover-based) magnonic excitation back to the low-intensity state \emph{without deactivating the pump} is not established. If clock-free operation is desired, the neuron should instead be operated below the bistable power window, where the neuron shows autonomous relaxation.

Finally, we comment on the perspective of an all-magnonic cascaded operation. While the present experiments employ electrical excitation and optical readout, the essential nonlinear interaction and the processing itself occur entirely within the magnonic system, which modulates the coupling to the applied RF pumps through the local magnon density. In particular, the cascade demonstrated in Fig.~\ref{fig:cascade}b operates with a common excitation frequency for all neurons, and with autonomous decay, the separate antennas can in principle be driven by a single, continuous RF source. More generally, one can envision architectures in which conversion back to another domain is only required in a final readout layer. Such readout could in principle be implemented electrically, for example using magnetoresistive detection methods such as GMR \cite{rossi.2025.magnetoresistivedetectionspin} or detection based on the spin Hall effect \cite{bracher.2017.detectionshortwavedspin}.

\subsection*{Functional benchmark of the observed activation function}\label{sec:benchmark}

The neuron exhibits rich temporal dynamics, which could enable memory and recurrent processing capabilities characteristic of spiking neural networks \cite{rodrigues.2023.spintronic_HH-Neuron, Hughes.2019.recurrentneural}. While such functionality is particularly relevant for processing time-dependent data, incorporating the full transient dynamics into larger network models is beyond the scope of the present work. Here, we instead consider a simplified, time-independent benchmark and assess whether the experimentally observed activation nonlinearity is already sufficient for standard learning tasks. For this purpose, we implement the neuron's activation function, derived from our numerical model, in a simplified neural-network model using the \textit{PyTorch} framework \cite{paszke.2019.pytorch}. We consider a compact convolutional neural network with sparse fan-in layers and restrict connectivity to remain compatible with small fully connected neuron blocks of at most $10 \times 10$ neurons per block. The detailed training setup, including the implementation of the differentiable activation interpolator and the network architecture, is described in Materials and Methods.

We evaluate image classification on the standard \textit{MNIST} \cite{mnist.2012.mnist} and \textit{fashionMNIST} \cite{xiao.2017.fashionmnist} datasets. As shown in Fig.~\ref{fig:neuralnetwork}a, despite the imposed constraints, the network achieves a classification accuracy of $\SI{97.15 \pm 0.049}{\percent}$ on \textit{MNIST} and $\SI{87.60 \pm 0.03}{\percent}$ on \textit{fashionMNIST}, averaged over $n=25$ runs with varying initializations. This indicates that the measured activation nonlinearity is principally suited for classification tasks.

A distinct feature of the magnonic neuron is that its activation function can be tuned via the pump power and frequency. Figure~\ref{fig:neuralnetwork}b shows the model-predicted activation landscape $\sigma(x, P_{\mathrm{Pump}}, f_{\mathrm{Pump}})$ and the corresponding differentiable interpolator used during training. In a scalable approach where a global RF pump is used, such tuning could alternatively be provided by local control of the neuron operating point, for example via voltage-controlled magnetoelectric coupling \cite{narducci.2024.magnetoelectriccouplingba} or via locally reconfigurable micro-magnets \cite{concconcelli.2024.nanomagnets, cocconcelli.2025.standaloneintegratedmagnonic}. Motivated by this tunability, we explore whether treating $P_{\mathrm{Pump}}$ and $f_{\mathrm{Pump}}$ as trainable parameters can improve performance under restricted model capacity. In large networks, abundant weights can easily compensate for trainable activation functions, resulting in a limited impact. To make such effects visible, we introduce an artificial bottleneck by reducing the number of neurons in the second-to-last layer from 8 to 4 (Methods). In this bottlenecked setting, trainable activations improve classification accuracy on \textit{fashionMNIST} from $\SI{76.14}{\percent}$ to $\SI{83.46}{\percent}$ (Fig.~\ref{fig:neuralnetwork}c) and enhance training robustness. After training, neuron pump powers systematically increase from their initial values, maximizing nonlinearity, while pump frequencies show moderate shifts, increasing the neuron threshold level (Fig.~\ref{fig:neuralnetwork}d). This suggests that trainable activation parameters could optimize performance under hardware constraints and also enable additional functionalities, such as dropout, compensation for device-to-device variations, or adaptation to fabrication tolerances.

While the present work focuses on the realization and understanding of the individual neuron mechanism and its connection to individual other neurons, it is instructive to outline a possible route toward interconnecting many neurons in future device concepts. One illustrative approach is to implement weighted connection layers via wave-based interference, for example using trainable/reconfigurable spin-wave scatterers as proposed in Ref.~\cite{papp.2021.spinwaveNN} (schematically shown in Fig.~\ref{fig:neuralnetwork}e). Realizing such interconnects will require addressing practical challenges including loss, cross-talk and phase stability, as well as practical means of training the array of magnets. 

A critical next step would therefore be the realization of a small-scale neural network using such devices. Beyond static classification, a particularly promising direction is to leverage the neuron's intrinsic temporal dynamics, where fading memory and history-dependent responses can provide an advantage for time-dependent processing tasks, for example in speech recognition or other temporal sequence problems \cite{larger.2017.highspeedphotonicreservoir}. For early usability studies, compact and resource-constrained on-device applications (''TinyML'') may be especially suitable, for instance in sensor-based classification, such as gas sensing and recognition using spiking or neuromorphic approaches \cite{huo.2023.bioinspiredspikingneural}.

\section*{Discussion}\label{sec:discussion}
We have experimentally realized an analog, all-magnonic neuron based on nonlinear magnon excitation in a Ga:YIG thin film. The neuron exhibits a tunable activation function driven by the intrinsic nonlinear magnonic frequency shift, enabling a sharp trigger response to external magnonic signals and tunable fading memory. Using a phenomenological flux model, we modeled these transient responses, providing an instructive understanding of the underlying mechanisms. Experimentally, we used these neuron properties to demonstrate multi-input neuron activation, leaky integration and ultimately cascading of three neurons by establishing synaptic connections via magnon propagation, providing key connectivity motifs for future magnonic neuromorphic concepts. Lastly, embedding the experimentally derived neuron activation function in a simplified neural-network model yielded high classification accuracy on standard benchmarks, supporting that the observed nonlinearity is sufficient for learning tasks and motivating future work toward interconnected magnonic architectures.

\section*{Materials and Methods}\label{sec:methods}
\subsection*{Film growth and sample fabrication}
The magnonic neuron system is based on a nanometer-thin gallium-substituted yttrium iron garnet (Ga:YIG) film. The film was grown on a gadolinium-gallium-garnet (111) substrate by liquid phase epitaxy (LPE) optimized for sub-100-nm single-crystalline \ce{Y3Fe_{5-x}Ga_{x}O12} films, following the procedures detailed in Ref. \cite{dubs.2025.gayig}, with a Ga-concentration of $x\sim 1.0$. The resulting film has a thickness of \SI{56}{\nano\meter} and exhibits a perpendicular magnetic anisotropy (PMA) of $\mu_{\mathrm{0}}H_{\mathrm{u}} = \SI{78.5}{\milli\tesla}$, a saturation magnetization of $\mu_{\mathrm{0}}M_{\mathrm{s}} = \SI{20.23}{\milli\tesla}$, and an exchange length of $\lambda_\text{ex} = \SI{90.37}{\nano\meter}$ which is significantly larger compared to standard YIG thin-films \cite{klingler.2015.exchangestiffness}. As a consequence, the magnon dispersion in Ga:YIG is strongly exchange-dominated, even at low wavevectors. This leads to a nearly isotropic and quadratic dispersion relation, which enables uniform and isotropic magnon propagation across the film, compare also Refs. \cite{breitbach.2024.erasing, boettcher.2022.gayig}. At the same time, the film shows ultra-low magnon damping, with a Gilbert damping parameter of $\alpha = \SI{5.59e-4}{}$. \textbf{Triangular structure}: Ground–signal–ground (GSG) CPW antennas were structured on top of the full film made from Ti/Au (\SI{10}{\nano\meter}/\SI{70}{\nano\meter}), deposited via electron beam evaporation and structured using electron beam lithography and lift-off processes. The respective width of the GSG lines is on average \SI{340}{\nano\meter}, with a G–S center-to-center distance of \SI{740}{\nano\meter}. \textbf{Waveguide structure}: YIG waveguides were fabricated using a hard mask ion beam milling procedure. This procedure was based on a Cr/Ti stack of \SI{30}{\nano\meter}/\SI{30}{\nano\meter} thickness as the hard mask and successive Ar+ ion milling under \SI{20}{\degree}, \SI{70}{\degree}, and \SI{20}{\degree} incident angles with respect to the film normal. In a final step, CPW antennas were structured on top of the waveguides made from Cr/Au (\SI{10}{\nano\meter}/\SI{100}{\nano\meter}) with similar methods as for the triangular structure.

\subsection*{BLS spectroscopy}
A single-mode laser with wavelength $\lambda = \SI{457}{\nano\meter}$ is focused onto the sample through the GGG substrate using a 100$\times$ magnification, compensating microscope objective (NA = 0.85), resulting in a spot size of approximately $\SI{300}{\nano\meter}$. The effective laser power on the sample is \SI{3}{\milli\watt}. The sample is mounted between the poles of an electromagnet to apply a homogeneous external magnetic field. Backscattered light from the sample is collected by the same objective and analyzed using a multipass tandem Fabry–Pérot interferometer connected to a single-photon counting detector, allowing for the detection of frequency shifts corresponding to magnon excitations due to conservation of energy and momentum \cite{sebastian.2015.uBLS}. The measured BLS intensity is proportional to the local magnon intensity and in-plane wavevectors of up to $k = \SI{24}{\radian\per\micro\meter}$ can be detected. Additionally, synchronization of the microwave excitation and detection is achieved using a pulse generator, enabling time-resolved BLS spectroscopy.

\subsection*{Calculation of the nonlinear frequency shift}
To calculate the nonlinear frequency shift and magnon dispersion relation in Ga:YIG, we use the Hamiltonian framework for nonlinear magnon theory \cite{patton.2010.hamiltonian}. This approach provides not only the linear dispersion relation, but also the coefficients for both nonlinear frequency self- and cross-shifts required to determine the modified dispersion at finite amplitudes.

We take into account the uniaxial anisotropy field $\mu_{\mathrm{0}} H_{\mathrm{u}} = {2 K_{\mathrm{u}}}/{M_{\mathrm{s}}}$ but disregard any effects from cubic anisotropy. Applying an external in-plane field $H_{\mathrm{ext}}$ we fully saturate the sample's magnetization in the in-plane direction, such that the internal effective magnetic field is $\mathbf{H}_{\mathrm{int}} = H_{\mathrm{x}}\hat{e}_{\mathrm{x}}$, and the effective magnetization is $M_{\mathrm{eff}} = M_{\mathrm{s}} - H_{\mathrm{u}}$. We then introduce the standard dipolar ``thin film function'' and  auxiliary function accounting for dynamic dipolar interaction and PMA:
\begin{equation}
f(x) = 1 - \frac{1-e^{-|x|}}{|x|}, \quad F_{\mathrm{zz,k}} = 1 - f(kd) - \frac{H_{\mathrm{u}}}{M_{\mathrm{s}}},
\end{equation}
together with $\omega_{\mathrm{M}} = \gamma \mu_{\mathrm{0}}M_{\mathrm{s}}$ and $\omega_{\mathrm{H}} = \gamma \mu_{\mathrm{0}}H_{\mathrm{int}}$, to define the intermediate expressions for backward-volume dipole-exchange magnons $\mathbf{k} \parallel \mathbf{M}$:
\begin{align}
Q_{\mathrm{k}}
  &= \frac{\omega_{\mathrm{M}}}{2}\Bigl(2\lambda^2 k^2 + F_{\mathrm{zz,k}}\Bigr), \notag\\
B_{\mathrm{k}}
  &= -\frac{\omega_{\mathrm{M}}}{2}F_{\mathrm{zz,k}}, \\
\Gamma_{\mathrm{zz,k}}
  &= \omega_{\mathrm{M}}\Bigl(\lambda^2 k^2 + f(kd)\Bigr). \notag
\end{align}
We have numerically verified that the dispersion and nonlinear shift results are almost exactly similar for the Damon-Eshbach geometry $\mathbf{k}\perp\mathbf{M}$ due to the high isotropy of the system. Hence, we disregard explicitly noting the formulas for other propagation directions at this point. Using $A_{\mathrm{k}} = \omega_{\mathrm{H}} + Q_{\mathrm{k}}$, the magnon eigenfrequency is given by $\omega_{\mathrm{k}} = \sqrt{A_{\mathrm{k}}^2 - |B_{\mathrm{k}}|^2}$. As referenced in the main text, the nonlinearly shifted magnon frequency of mode $k$, influenced by the nonlinear self-shift by its own amplitude $c_{\mathrm{k}}$, and cross-shift by the magnon mode $k'$ with amplitude $c_{\mathrm{k'}}$ is computed via Eq. (\ref{equ:nonlinear_shift}). The self-shift coefficient is explicitly given by:
\begin{equation}
    T_{\mathrm{k}} = -Q_{\mathrm{k}} + \frac{B_{\mathrm{k}}^2}{2\omega_{\mathrm{k}}^2}\left(\omega_{\mathrm{H}} + \Gamma_{\mathrm{zz,2k}}\right) + \left(1+\frac{B_{\mathrm{k}}^2}{\omega_{\mathrm{k}}^2}\right)\Gamma_{\mathrm{zz,0}}.
\end{equation}
The more elaborate nonlinear cross-shift coefficient, using shorthand notation $1 \equiv k$ and $2 \equiv k'$, reads:
\begin{align}
&T_{\mathrm{12}} = W_{\mathrm{12,12}} = \Psi_{\mathrm{12,(-1)(-2)}}\left(u_1^2 u_2^2 + v_1^2 v_2^2\right) + \notag \\
&\Psi_{\mathrm{2(-1),1(-2)}}\left(u_1^2 v_2^2 + v_1^2 u_2^2\right) + 
2 \Psi_{\mathrm{1(-1),2(-2)}} u_1 u_2 v_1 v_2 + \notag \\
&2 \Phi_{\mathrm{112,2}} u_1 v_1 \left(u_2^2 + v_2^2\right) + 
2 \Phi_{\mathrm{122,1}} u_2 v_2 \left(u_1^2 + v_1^2\right). 
\end{align}
The terms above rely on four-wave mixing coefficients defined over general combinations of wavevectors $k_{\mathrm{1}}$ through $k_{\mathrm{4}}$, denoted as $\mathrm{1},\mathrm{2},\mathrm{3},\mathrm{4}$:
\begin{align}
\Psi_{\mathrm{12,34}} &= -\frac{1}{4} \sum_{i=1}^{4} Q_i + \frac{1}{4} \sum_{i=1}^{2} \sum_{j=3}^{4} \Gamma_{\mathrm{zz}, i+j}, \notag \\
\Phi_{\mathrm{123,4}} &= -\frac{1}{4} \sum_{i=1}^{3} B_i, \\
u_{\mathrm{k}} &= \sqrt{\frac{A_{\mathrm{k}} + \omega_{\mathrm{k}}}{2 \omega_{\mathrm{k}}}}, \quad 
v_{\mathrm{k}} = -\frac{B_{\mathrm{k}}}{|B_{\mathrm{k}}|} \sqrt{\frac{A_{\mathrm{k}} - \omega_{\mathrm{k}}}{2 \omega_{\mathrm{k}}}}. \notag
\end{align}
Here, we used the fact that $B_{\mathrm{k}}$ is real in our case.

\subsection*{Numerical neuron model}
We describe the magnonic neuron as a single effective mode with magnon intensity $|\upr{c}{N}(t)|^2$. The model captures the feedback loop that underlies the observed threshold behavior: a nonlinear frequency shift changes the excited wavevector $k$, which in turn modulates the CPW excitation efficiency $\eta(k)$, thereby altering the inflow of magnons. The temporal evolution of the neuron intensity is described by
\begin{equation}\label{equ:ODE2}
\begin{split}
\frac{\mathrm{d}|\upr{c}{N}(t)|^2}{\mathrm{d}t}
    &= \eta(k[|\upr{c}{total}(t)|^2])\upr{P}{MW} \\
    &\quad - \lambda|\upr{c}{N}(t)|^2
      - \lambda'|\upr{c}{N}(t)|^4.
\end{split}
\end{equation}
where both, the neuron intensity $|\upr{c}{N}(t)|^2$ and the input intensity from other neurons $|\upr{c}{in}(t)|^2$ contribute to the nonlinear shift, here represented by $|\upr{c}{total}(t)|^2$. The external input intensity $|\upr{c}{in}(t)|^2$ was implemented as a nanosecond-short Gaussian pulse to match the measured trigger waveforms. Note that in our model, the trigger activates the neuron passively by changing the pump efficiency, but does not directly contribute to the intensity of the neuron mode $|\upr{c}{N}(t)|^2$. This is because the trigger runs on a different mode, which propagates. We solve this equation in a discrete time-domain simulation using Euler's method ($\Delta t_\mathrm{sim}=\SI{0.1}{ns}$), i.e., for each timestep $t$ we 
\begin{enumerate}[label=(\roman*), itemsep=0pt, topsep=0.5em]
    \item Calculate the instantaneous wavevector $k(t,\omega[\upr{c}{total}(t)])$ from the nonlinear dispersion using the total magnon intensity
    \item Compute the antenna excitation efficiency $\eta(k(t))$
    \item Calculate the change in neuron intensity following Eq. (\ref{equ:ODE2})
    \item Add random amplitude noise by replacing $|\upr{c}{N}(t)|^2 \rightarrow \big(\sqrt{|\upr{c}{N}(t)|^2} + \xi_t\big)^2$, with $\xi_t \sim \mathcal{N}\!\big(\mu = 0,\sigma_\mathrm{amp}^2\,\Delta t_\mathrm{sim}\big)$ (zero-mean Gaussian distribution; the $\sqrt{\Delta t_\mathrm{sim}}$ scaling ensures the noise power stays constant when changing the simulation timestep $\Delta \upr{t}{sim}$)
    \item Determine the intensity for the next timestep 
    $|\upr{c}{N}(t+\Delta t)|^2 =|\upr{c}{N}(t)|^2 + \Delta t \frac{\mathrm{d}|\upr{c}{N}(t)|^2}{\mathrm{d}t}$.
\end{enumerate}
The fixed simulation parameters used for the numerical model are listed in Table \ref{tab:sim-params}.

\begin{table}[h!]
\caption{Simulation parameters used in the numerical neuron model of Eq.~\eqref{equ:ODE2}. Frequencies are given as $f=\omega/2\pi$.}
\label{tab:sim-params}
\centering
\sisetup{mode=text,detect-weight=true,table-number-alignment=center}
\begin{tabular*}{\columnwidth}{@{\extracolsep{\fill}} l l}
\toprule
\textbf{Parameter (unit)} & \textbf{Value} \\
\midrule
Self shift coeff.: $T_{k}$ (\si{\giga\hertz}) & \num{1.89} \\
Cross shift coeff.: $T_{kk'}$ (\si{\giga\hertz}) & \num{1.57} \\
\shortstack[l]{Quadratic Dispersion\\ Coeff.: $D$ (\si{\micro \meter^2.\mega\hertz.\radian^{-2}})} & \num{4.91} \\

FMR freq.: $f_{\mathrm{FMR}}$ (\si{\giga\hertz}) & \num{1.39} \\
Pump freq.: $f_{\mathrm{exc}}$ (\si{\giga\hertz}) & \num{1.69} \\
Sim. timestep: $\Delta t_{\mathrm{sim}}$ (\si{\nano\second}) & \num{0.1} \\
Pulse dur.: $t_{\mathrm{pulse}}$ (\si{\nano\second}) & \num{10} \\
Linear damping: $\lambda$ (\si{\nano\second^{-1}}) & \num{0.0167} \\
Nonlinear damping: $\lambda'$ (\si{\nano\second^{-1}}) & \num{0.5} \\
Noise int.: $\sigma_{\mathrm{amp}}^2$ (a.u.) & \num{0.01} \\
Averages: $n_{\mathrm{avg}}$ (—) & \num{1e4} \\
\botrule
\end{tabular*}
\end{table}

\subsubsection*{Wavevector calculation}
To calculate the instantaneous wavevector $k(\omega,t)$, we invert the nonlinearly shifted dispersion function. With regard to numerical efficiency, we approximate the dispersion for linear excitation, which is almost perfectly quadratic, with the expression:
\begin{equation}
    \omega(k,t) = \upr{\omega}{FMR} + Dk^2 + \Delta\upr{\omega}{NL}(\upr{c}{total},t).
\end{equation}
The FMR frequency $\upr{\omega}{FMR}$ as well as the quadratic coefficient $D$ are obtained by a fit to the linear dispersion relation calculated by $\upr{\omega}{k} = \sqrt{\upr{A}{k}^2 - |\upr{B}{k}|^2}$ in the wavevector regime of interest $k<\SI{20}{\radian\per\micro\meter}$, yielding an almost perfect fit with $R^2 = 0.9995$. The nonlinear frequency shift is driven via nonlinear self-shift $\upr{T}{k}$ by the neuron itself, and via cross-shift $\upr{T}{kk'}$ by an external input, provided the external input runs on a different magnon mode. The resulting shift is given by 
\begin{equation}
    \Delta\upr{\omega}{NL}(k,t) = \upr{T}{k}|\upr{c}{N}(t)|^2 + \upr{T}{kk'}|\upr{c}{in}(t)|^2,
\end{equation}
and the instantaneous wavevector is obtained by
\begin{equation}
k^2(t) = {\frac{\upr{\omega}{k} - \upr{\omega}{FMR} - \upr{T}{k}|\upr{c}{N}(t)|^2 + \upr{T}{kk'}|\upr{c}{in}(t)|^2}{D}}.
\end{equation}

\subsubsection*{Excitation efficiency}
Efficient excitation is obtained when the spatial periodicity of the antenna field matches the periodicity of the magnon being excited. For \textit{linear} excitation, i.e., at low amplitudes, the excitation efficiency $\eta(k)$ is proportional to the spatial Fourier transform of the antenna's \O ersted field \cite{schneider.2008.efficiency, demidov.2009.efficiency}. We model the nonlinear excitation efficiency as follows. We define the critical amplitude $\upr{c}{crit}$ for which the excitation efficiency reaches its maximum $\eta(k[\upr{c}{crit}]) = 1$.
\begin{enumerate}[label=(\roman*), itemsep=0pt, topsep=0.5em]
    \item For small amplitudes $c<\upr{c}{crit}$, we calculate the linear excitation efficiency following the procedure above
    \item For large amplitudes $c\geqslant\upr{c}{crit}$, we fix $\eta(k[c]) \equiv 1$.
\end{enumerate}
Hence, our model captures the gain in excitation efficiency at small amplitudes, accounting for the positive feedback mechanism underlying the neuron's threshold behavior, but experiences no drastic additional gain or loss for larger amplitudes, which aligns with experimental observations of foldover systems.

\subsubsection*{Damping}
The magnon lifetime is given by $\tau = (2\pi\alpha\upr{A}{k})^{-1}$, with the Gilbert damping parameter $\alpha$ and $\upr{A}{k}$ as introduced above. The linear damping $\lambda$ was estimated from the inverse lifetime $\lambda = 2\tau^{-1}$ (factor 2 for intensity) at the wavevector of initial excitation. The nonlinear damping parameter $\lambda'$ was obtained empirically by matching the simulated intensity–power characteristics to the BLS data, selecting the value that yielded the best overall agreement.

\subsubsection*{Comparison to BLS data and limitations}
The measured BLS intensity is proportional to the magnon intensity but scales with an unknown sensitivity prefactor as well as an offset induced by noise and dark count rate. Hence, for overlays with BLS data, simulated intensities were mapped by a linear transformation $I_\mathrm{BLS}=\beta\,|\upr{c}{N}|^2+\beta_0$, with $\beta,\beta_0$ obtained by matching baseline and peak levels. Similarly, the magnon intensity driven by the antenna excitation is proportional to $\eta\upr{P}{MW}$, but scales with an unknown prefactor $\alpha$, as an unknown fraction of the microwave power is absorbed by the magnonic system. Hence, the simulated powers were mapped by a linear transformation without offset to the experimental powers $\upr{P}{sim} = \alpha \upr{P}{exp}$.

\subsubsection*{Limitations}
The model is single-mode and intensity-only, and does not account for phase or spatial dependencies. These approximations are sufficient to reproduce the observed threshold, saturation, and tunable decay, but they do not capture any higher complexity, such as multi-mode or spatial interference.

\subsection*{PyTorch neural network training}

To benchmark the usefulness of the neurons nonlinearity, we implemented the activation function predicted by our model as a custom activation function into the PyTorch framework \cite{paszke.2019.pytorch}. We first computed the activation function $\sigma(x, P_{\mathrm{Pump}}, f_{\mathrm{Pump}})$ on a 3D grid of the input intensity, pump power $P_{\mathrm{Pump}}$, and pump frequency $f_{\mathrm{Pump}}$ of the neuron. We used this tensor to construct a fast and differentiable interpolator using PyTorch's \texttt{grid\_sample} function with bilinear interpolation. The additional parameters $P_{\mathrm{Pump}}$ and $f_{\mathrm{Pump}}$ were mapped to an unconstrained form using a sigmoid function, ensuring smoothness and learning within the parameter bounds. The activation function was furthermore extended to take the square of the input, $c_{\mathrm{out}} = \sqrt{\sigma(c_{\mathrm{in}}^2)}$. This treatment leaves the network computation in the amplitude domain while mimicking the neuron’s intensity sensitivity, allowing activation from both positive and negative amplitudes.

The network architecture consisted of a fixed preprocessing pipeline, followed by sparsely connected fan-in layers. The preprocessing stage included two convolutional layers intended as linear feature extractors suitable for passive wave-based or FPGA-based implementations (kernel size 3, padding 1; channel progression: 1~$\rightarrow$~32~$\rightarrow$~64), each followed by a single 2$\times$2 max pooling kernel to reduce the spatial resolution from $28 \times 28$ to $7 \times 7$. If strict physical consistency is required, this operation could be replaced by an average pooling layer. 

The output of preprocessing was passed through three sparsely connected neural network layers with dimensionality reductions of 3136~$\rightarrow$~448, 448~$\rightarrow$~64, and 64~$\rightarrow$~8, with block-wise connection ratios of 7:1, 7:1, and 8:1, ensuring sparse connectivity throughout the classification layers and remaining within the hardware constraint of at most $10 \times 10$ fully connected neurons. The final classification layer mapped 8~$\rightarrow$~10 output classes. The custom neuron activation function was applied in all fan-in and output layers. For the bottlenecked variant of the network, we exceeded this hardware constraint by reducing the second-to-last layer from 8 to 4 neurons, resulting in a 16:1 connection ratio. This configuration was intentionally chosen to restrict the model capacity and highlight how the additional degrees of freedom provided by trainable neuron parameters can improve performance when the network capacity is limited.

All networks were trained using the Adam optimizer for 100 epochs with a batch size of 64. An adaptive learning rate scheduler reduced the learning rate on validation plateaus. Each experiment was repeated across 25 random seeds, and the same seed set was used for all architectures and dataset variants to ensure statistical comparability across configurations.

\section*{Supplementary materials}
\textbf{This PDF file includes:}\\
Fig. S1 - Nonlinear magnonic frequency shift\\
Fig. S2 - Numerical neuron model\\
Fig. S3 - Magnon lifetime and decay length\\
Fig. S4 - Fading memory\\
Fig. S5 - Intrinsic signal accumulation\\
Fig. S6 - Waveguide neuron bistable foldover\\
Fig. S7 - Non-reciprocal neuron output\\

\nocite{heinz.2021.longrangespinwavepropagation}


%

\section*{Acknowledgements, funding}

This research was funded by the European Research Council within the Starting Grant No. 101042439 "CoSpiN", by the Deutsche Forschungsgemeinschaft (DFG, German Research Foundation) within the Transregional Collaborative Research Center—TRR 173–268565370 “Spin + X” (project B01), and the project 271741898. DB acknowledges support by the Max Planck Graduate Center with the Johannes Gutenberg-Universität Mainz (MPGC). MC, GF and DR are with Petaspin team and thank the support of the PETASPIN association (www.petaspin.com). Q.W. acknowledges support within the National Natural Science Foundation of China (Grant No. 12574118). RV acknowledges support by the NAS of Ukraine, Project No 0124U000270.

\section*{Author contributions}
\begin{itemize}[itemsep=1mm, parsep=0pt]
    \item[--] Conceptualization: DB, MB, HM, BH, DR, AAH, PP
    \item[--] Material contribution: CD
    \item[--] Sample preparation: MB
    \item[--] Investigation: DB
    \item[--] Visualization: DB
    \item[--] Discussion: BH, RV, QW, MC, GF, DR, AAH, PP
    \item[--] Supervision: PP
    \item[--] Writing, original draft: DB
    \item[--] Writing, review \& editing: DB, MB, BH, RV, QW, CD, MC, GF, DR, AAH, PP
\end{itemize}

\section*{Competing interests}
The authors declare that they have no competing interests.

\section*{Data and materials availability}
All data needed to evaluate the conclusions in the paper are present in the paper and/or the Supplementary Materials. The codes that support the theoretical modelling and calculations are available on Zenodo at \href{https://doi.org/10.5281/zenodo.18657410}{10.5281/zenodo.18657410} and \href{https://doi.org/10.5281/zenodo.18657424}{10.5281/zenodo.18657424}.

\end{document}